\documentclass[sigconf,nonacm,balance=false]{acmart}

\usepackage{popets}
\setcopyright{popets}
\copyrightyear{YYYY}
\acmYear{YYYY}
\acmVolume{YYYY}
\acmNumber{X}
\acmDOI{XXXXXXX.XXXXXXX}
\acmISBN{}
\acmConference{Proceedings on Privacy Enhancing Technologies}
\usepackage{graphicx} 

\usepackage{longtable}
\usepackage{ragged2e}
\usepackage{makecell}

\usepackage{multirow}
\usepackage{amsmath,trimclip}
\usepackage{rotating} 
\usepackage{mdframed}
\usepackage{url}
\usepackage{yfonts} 
\usepackage{enumitem}
\usepackage{xspace}

\makeatletter
\DeclareRobustCommand{\circbullet}{\mathbin{\vphantom{\circ}\text{\circbullet@}}}
\newcommand{\circbullet@}{%
  \check@mathfonts
  \m@th\ooalign{%
    \clipbox{0 0 0 {\dimexpr\height-\fontdimen22\textfont2}}{$\bullet$}\cr
    $\circ$\cr
  }%
}
\makeatother

\newcommand{\opencircle}{\ensuremath{\circ}}
\newcommand{\filledcircle}{\ensuremath{\bullet}}
\newcommand{\halffilledcircle}{\ensuremath{\circbullet}}

\newcommand{\0}{\opencircle}
\newcommand{\1}{\halffilledcircle}
\newcommand{\2}{\filledcircle}

\newcommand{\SPtech}{\textbf{\textsf{Tech}}}
\newcommand{\SPtechDP}{\textbf{\textsf{TechDP}}}
\newcommand{\SPend}{\textbf{\textsf{End}}}
\newcommand{\SIsurvey}{\textbf{\textsf{SURV}}}
\newcommand{\SIfocusGroup}{\textbf{\textsf{FOC}}}
\newcommand{\SItaskCompletion}{\textbf{\textsf{TASK}}}
\newcommand{\SIinterview}{\textbf{\textsf{INT}}}
\newcommand{\EDbetweenSubjects}{\textbf{\textsf{B}}}

\renewcommand{\EDbetweenSubjects}{\textbf{EB}}

\newcommand{\EDwithinSubjects}{\textbf{\textsf{W}}}
\renewcommand{\EDwithinSubjects}{\textbf{EW}}

\newcommand{\ReAMT}{\textbf{\textsf{AMT}}}
\newcommand{\ReProlific}{\textbf{\textsf{Pro}}}
\newcommand{\ReQualtrics}{\textbf{\textsf{Q}}}
\newcommand{\ReNetwork}{\textbf{\textsf{Net}}}
\newcommand{\ReOutreach}{\textbf{\textsf{Out}}}
\newcommand{\ReOutsource}{\textbf{\textsf{Con}}}
\newcommand{\GeoUS}{\textbf{\textsf{US}}}
\newcommand{\GeoEurope}{\textbf{\textsf{EU}}}

\newcommand{\numpapers}[0]{27\xspace}

\title{SoK: Usability Studies in Differential Privacy}

\newcommand{\spheading}[2][70pt]{
  \rotatebox{90}{\parbox{#1}{\raggedright #2}}}

\renewcommand{\paragraph}[1]{\vspace*{1pt plus 1pt minus 1pt}\noindent\textbf{#1}\hspace*{4pt}}
\newcommand{\iparagraph}[1]{\subsubsection{#1}}

\author{Onyinye Dibia}
\authornote{These authors contributed equally to this work.}
\email{Onyinye.Dibia@uvm.edu}
\affiliation{\institution{University of Vermont} \country{}}

\author{Prianka Bhattacharjee}
\authornotemark[1]
\email{Prianka.Bhattacharjee@uvm.edu}
\affiliation{\institution{University of Vermont} \country{}}

\author{Brad Stenger}
\authornotemark[1]
\email{jstenger@uvm.edu}
\affiliation{\institution{University of Vermont} \country{}}

\author{Steven Baldasty}
\email{Steven.Baldasty@uvm.edu}
\affiliation{\institution{University of Vermont} \country{}}

\author{Mako Bates}
\email{Mako.Bates@uvm.edu}
\affiliation{\institution{University of Vermont} \country{}}

\author{Ivoline C. Ngong}
\email{kngongiv@uvm.edu}
\affiliation{\institution{University of Vermont} \country{}}

\author{Yuanyuan Feng}
\email{Yuanyuan.Feng@uvm.edu}
\affiliation{\institution{University of Vermont} \country{}}

\author{Joseph P. Near}
\email{jnear@uvm.edu}
\affiliation{\institution{University of Vermont} \country{}}

\begin{document}

\begin{abstract}
Differential Privacy (DP) has emerged as a pivotal approach for safeguarding individual privacy in data analysis, yet its practical adoption is often hindered by challenges in the implementation and communication of DP. This paper presents a comprehensive systematization of existing research studies around the usability of DP, synthesizing insights from studies on both the practical use of DP tools and strategies for conveying DP parameters that determine privacy protection levels, such as epsilon($\varepsilon$). By reviewing and analyzing these studies, we identify core usability challenges, best practices, and critical gaps in current DP tools that affect adoption across diverse user groups, including developers, data analysts, and non-technical stakeholders. Our analysis highlights actionable insights and pathways for future research that emphasizes user-centered design and clear communication, fostering the development of more accessible DP tools that meet practical needs and support broader adoption.

\end{abstract}

\maketitle

\section{Introduction}
\label{sec:introduction}
In an era where data is increasingly viewed as a vital asset, the need for robust privacy protections has never been more critical. Differential privacy (DP) has emerged as a cornerstone of modern data privacy, offering robust privacy guarantees that allow for meaningful data analysis while protecting the privacy of individuals~\cite{dwork2006calibrating, dwork2014algorithmic}. Government agencies, large corporations, and research institutions have integrated DP into various data practices ~\cite{nanayakkara2023chances, karegar2022exploring}, but its real-world adoption among small- to med-sized organizations and  its broader acceptance among the public remain limited.

The inherent complexity of DP leads to barriers for real-world implementation and user understanding.
First, the effective implementation of DP requires adequate understanding of its underlying mechanisms, such as the concept of privacy budget ($\varepsilon$) that determines the level of privacy protection~\cite{nanayakkara2023chances}. The implementation often requires specialized DP tools, which can be difficult to use even for data practitioners with technical backgrounds~\cite{ngong2024evaluating}. 
Second, it is hard to accurately communicate DP's privacy protection levels, which can prevent the general public and other stakeholders from fully understanding the privacy implications of a given DP implementation. Failure to convey DP’s technical details may lead to misconceptions and decreased acceptance~\cite{xiong2022using, xiong2023exploring}.

These barriers underscore the challenges around \textbf{DP's usability}, which we define in this systemization of knowledge (SoK) paper as \textbf{the ease and effectiveness that different users engage with DP in ways that suit them.} Specifically, this is a two-pronged definition of usability considering the needs of two distinct user groups: \textbf{technical users} (e.g., data practitioners, software developers) who often need to use DP tools to implement DP in their systems, where we adopt the classic usability definition for software (e.g., learnability, efficiency, error prevention)~\cite{nielsen1994measuring} to consider DP tools' usability; and \textbf{end users} (e.g., data subjects or the general public) of DP implementations, where we consider usability as the effectiveness of communicating the privacy implications of DP, since this is the primary way end users engage with DP.


To navigate the challenges around DP's usability, this paper aims to synthesize and critically evaluate existing research on both the usability of DP tools~\cite{ngong2024evaluating, gaboardi2018psi, sarathy2023don,murtagh2018usable, marcusland2020, 9629406, alrige2013understanding} and strategies for communicating DP concepts effectively to diverse audiences~\cite{nanayakkara2023chances,cummings2021need, franzen2022private, karegar2022exploring, nanayakkara2022visualizing, smart2022understanding, kuhtreiber2022replication, xiong2022using, xiong2020towards, xiong2023exploring, wen2023influence, bullek2017towards}. By systematically reviewing the literature, we explore two central aspects of DP usability: the creation and selection of DP tools, and the communication of DP-specific parameters. Usability studies, in this context, refer to empirical research studies that examine not only the ease of use of DP tools but also the methods for effectively communicating about DP guarantees.
Our work aims to provide researchers with an overview of usability studies in DP and to highlight important directions for future work.


In summary, we make the following contributions:
\begin{itemize}[leftmargin=14pt, itemsep=0pt, topsep=1pt]
    \item \textbf{Synthesizing existing usability studies}: We review and summarize findings from \numpapers studies on DP tool usability and communication strategies, analyzing their methodologies, target audiences, and study outcomes.
    \item \textbf{Identification of Best Practices}: We identify best practices for conducting usability studies in DP, with the objective of improving the design and evaluation of DP tools, as well as the communication strategies used to convey DP concepts.

    \item \textbf{Highlighting Research Gaps}:  We highlight open research challenges in DP usability, providing a roadmap for future research. These challenges include the need for more standardized communication methods, the development of more user-friendly tools for technical users without DP expertise, and the integration of usability considerations into implementations.
\end{itemize}

\paragraph{Overview of this paper.}
The technological development of DP is on a trajectory where usability increasingly influences the decisions made by DP adopters. Our analysis framework divides DP usability into two main strata, "DP tools" and "DP communication", each with its own set of comparison points and discussion themes. 

Section~\ref{sec:background} summarizes the basics of DP and recent relevant research.
Section~\ref{sec:our-methods} describes our process for finding and choosing papers to include in this SoK work.
Section~\ref{sec:methods-of-usability} presents our analysis of the methodologies employed these usability studies related to both DP tools and DP communication. 
Section~\ref{sec:results_usability} delves into the findings of usability studies centered on DP tools and DP communications, separately.
The first part of the section summarizes how DP software tools designed are assessed; the second part of this section reviews usability studies around DP communication.
%
Section~\ref{sec:discussion} identifies specific takeaways and challenges that emerged from our review, along with our recommendations and proposed solutions; Section~\ref{sec:open_questions} discusses important open questions for future research.

\section{Background \& Related Work}
\label{sec:background}

\paragraph{Differential privacy (DP).}
Differential Privacy (DP) \cite{dwork2006calibrating, dwork2014algorithmic} is a formal privacy definition designed to enable the statistical analysis of data while safeguarding individual privacy. At its core, DP introduces randomness, typically in the form of noise, into the analysis process to obscure the impact of any single data point, thereby preventing the identification of individuals within a dataset. Formally, two datasets \( D \) and \( D' \) are considered neighboring datasets if they differ by only one individual's data. A mechanism \( M \) satisfies \((\varepsilon, \delta)\)-DP if, for all neighboring datasets \( D \) and \( D' \) and for all possible outcomes \( S \), $\Pr[M(D) \in S] \leq e^{\varepsilon} \cdot \Pr[M(D') \in S] + \delta$.
Here, epsilon ($\varepsilon$) represents the privacy parameter or privacy budget, where a smaller $\varepsilon$ value indicates stronger privacy guarantees, while a larger $\varepsilon$ value indicates weaker privacy, and delta (\( \delta \)) is a small positive value that accounts for unlikely outcomes.
(\(\delta\)) is set to zero for \emph{"pure"} DP.
Noise, often drawn from distributions such as the Laplace or Gaussian, is used to achieve differential privacy~\cite{dwork2006calibrating, dwork2014algorithmic}.

\paragraph{SoK work in DP.} Mainstream DP research has examined various aspects of the design of DP algorithms, covering the theory of DP~\cite{dwork2014algorithmic}, various DP definitions~\cite{desfontaines2020sok}, and interpretations of the guarantee~\cite{tschantz2020sok}. 
Previous DP SoK work focused on specific algorithms for histograms~\cite{nelson2019sok}, database queries~\cite{near2021differential}, mobility trajectories~\cite{miranda2023sok}, and graphs~\cite{mueller2022sok}.
More recently, the DP research community has begun to acknowledge the importance of making DP more usable to different stakeholders~\cite{cummings2023advancing}. 
Cummings and Sarathy~\cite{cummings2023centering} further identified research gaps around usable DP, such as the need to tailor DP communication for different audiences and the lack of clear guidance for interpreting privacy-loss parameters ($\varepsilon$, $\delta$).

However, there is no SoK work on existing efforts to make DP more usable, which is essential for broadening the real-world adoption of DP. 
Therefore, we set out to perform a systematic review of existing research on the usability of DP, which broadly refers to studies at the intersection of DP and human-computer interaction that involve usability evaluations or user studies. In our SoK, we refer to this body of work as ``DP usability studies.'' Our goal is to summarize results in this area for researchers in the DP and human-computer interaction research communities, and to highlight important open questions for future research.

\paragraph{DP \& the law.}
A significant body of work in the law community has examined the connection between formal privacy definitions like DP and legal requirements. Solove~\cite{solove2005taxonomy} presents a taxonomy of privacy informed by laws and regulations; the taxonomy does not consider formal privacy definitions. Solove's taxonomy is based on a deep analysis of legal opinions, and is one starting point for subsequent work connecting formal privacy definitions like DP to legal and regulatory requirements.
Nissim and Wood~\cite{nissim2018privacy} analyze the gaps between \emph{normative} definitions of privacy (including laws and regulations) and \emph{technical} definitions of privacy (including differential privacy). They identify significant gaps between the two, and recommend both improving privacy regulations and considering additional layers of technical definition (e.g. contextual integrity~\cite{nissenbaum2004privacy}) to bridge these gaps.
Sokolovska and Kocarev~\cite{sokolovska2018integrating} develop a legal and a technical framework for privacy, then analyze the differences. Like Nissim and Wood, they find that additional layers of formalism may be needed to align with legal requirements.

GDPR compliance is particularly well-studied. Holzel~\cite{holzel2019differential} examines the specific question of how DP relates to GDPR's anonymization requirement, and identifies several important gaps between the GDPR's requirements and the standard central-model approach for deploying DP. Cohen and Nissim~\cite{cohen2020towards} examine the specific legal concept of ``singling out'' in the GDPR. They provide a formal definition of this concept, and prove that DP ensures protection against singling out.

Additional work has focused on specific deployments as case studies for analyzing legal issues associated with DP.
Chin and Klienfelter~\cite{chin2011differential} analyze the connection between DP and the law via the case study of Facebook's advertising interface. They identify specific properties of public-facing deployments that make DP an appropriate choice for legal compliance, based on Solove's taxonomy and their own analysis of legal requirements.
Cohen et al.~\cite{cohen2022private} analyze the specific legal requirements associated with the Decennial Census, and find that the DP-based approach adopted by the US Census Bureau generally satisfies them. These requirements include both privacy and certain specific utility requirements (e.g. for balancing electoral districts).

\section{Review Procedure}
\label{sec:our-methods}

We identified \numpapers research papers from a literature search across four digital libraries. Here, we detailed our review procedure. 

\paragraph{Paper Selection.}
To summarize the research development around the usability of Differential Privacy (DP), we conducted a structured literature search across four digital libraries: Google Scholar, ACM Digital Library, IEEE Xplore, and Semantic Scholar. The search was performed on July 25, 2024, and targeted papers published after 2010. We used the following search string: \texttt{"differential privacy" AND ("usable" OR "usability") AND "participants"}. This combination was chosen to capture empirical studies that explore human-centered aspects of DP—specifically, studies that involved participants and addressed usability either in tool interaction or in communication of privacy guarantees.
Some relevant studies may use alternate terminology such as "comprehension," "perceptions," or "accessibility." Our manual review of alternative search strings including these terms did not reveal additional relevant papers missed by our original search terms.

Our inclusion criteria required that papers (1) conduct a study with human subjects (2) related to the ease and effectiveness that different users engage with DP. 
%
We compiled the search results into a codebook (as a spreadsheet) containing 757 potentially relevant papers for manual review. Since Google Scholar does not expose an API, we manually scraped the first 200 results. Because Semantic Scholar ranks results by relevance, we reviewed the first 200 entries from that source as well. IEEE and ACM returned fewer than 200 papers in response to the search query.




\paragraph{Codebook development.} We developed a structured codebook through a collaborative, iterative process involving all seven authors over the course of several months.
%
We began by dividing the initial list of 757 potentially relevant papers among the authors. Each author reviewed abstracts and applied predefined inclusion criteria: papers had to either conduct a user study or propose a software or methodological tool aimed at improving usability in DP. We defined usability as encompassing one or more of the following: end-user understanding of DP concepts, curator ability to set appropriate privacy levels, and analyst ability to leverage data. After the initial individual round of review, the authors met to finalize inclusion decisions, resolve disagreements, eliminate duplicates, and arrive at the final list of 27 papers for in-depth analysis.

We completed the codebook for the remaining 27 papers by deductive coding~\cite{bingham2021deductive} informed by prior literature and existing frameworks on evaluating usability. We iteratively refined the codes through team discussions. The final codebook included categories for both methodology and study findings.
%
%
Each paper was double-coded by at least two authors. In cases of discrepancy, the authors discussed their reasoning during weekly team meetings and reached consensus through group discussion. These meetings also served to refine the interpretation of codes and resolve ambiguities. For example, we aligned on how to distinguish between “objective” and “subjective” understanding and standardized how we annotated outcomes like satisfaction and willingness to share.



\paragraph{Limitations.}
%
Our review of relevant papers relied on selecting the most appropriate keywords, and could have inadvertently excluded some papers. Our use of specific databases (and the omission of certain libraries like DBLP) could also have contributed to the omission of relevant research. Furthermore, the scope of this review may be limited by its focus on English-language publications, potentially overlooking valuable contributions in other languages. Finally, usability in DP is an active area of research, and our review may miss papers released after our review.

\section{Methodology of DP Usability Studies}
\label{sec:methods-of-usability}

We analyzed and categorized the methodologies used in the usability studies reviewed in our SoK. Key area covered include studies' recruitment, study design and instruments, and evaluation criteria that measure various aspects of DP tool's usability or user's DP understanding.
Tables~\ref{tab:methodology_Studies_Investigating_Software_Tools} and~\ref{tab:methodology_communicating_dp} summarize our analysis results of the \numpapers papers on DP software tools and DP communication, respectively.

\begin{table*}[tbhp]
\centering
\begin{tabular}{|l|p{50pt}|l|p{100pt}|l|l|l|l|l|l|l|l|l|}
\hline
\multirow{2}{*}{\spheading{\textbf{Citation}}} & 
\multirow{2}{*}{\spheading{\textbf{Recruitment}}} & 
\multirow{2}{*}{\spheading{\textbf{Geo-location}}} & 
\multirow{2}{*}{\spheading{\textbf{Sample Populations}}} & 
\multirow{2}{*}{\spheading{\textbf{Sample Size}}} & 
\multirow{2}{*}{\spheading{\textbf{Study Methods \\ \& Instruments}}} & 
\multirow{2}{*}{\spheading{\textbf{Educational Materials}}} &
\multicolumn{6}{|c|}{\textbf{Evaluation Methods}} \\ \cline{8-13}
&&&&&&&
\spheading{\textbf{Objective Understanding}} & 
\spheading{\textbf{Subjective Understanding}} & 
\spheading{\textbf{Task Success Rate}} & 
\spheading{\textbf{Time on Task}} & 
\spheading{\textbf{Error Rate}} & 
\spheading{\textbf{Satisfaction}} \\ \hline

\cite{9629406} & \ReNetwork & $\GeoEurope$ & \SPtech & 8 & \SIfocusGroup & \checkmark & {\1}&{\2} & {\0} & {\0} & {\0} & {\2}  \\ \hline 
\cite{ngong2024evaluating} & \ReNetwork & $\GeoUS$ &  \SPtechDP & 24 & \SIsurvey, \SIinterview, \SItaskCompletion, $\EDbetweenSubjects$  & \checkmark & {\2} & {\1}&{\2} & {\2} & {\2} & {\2}   \\ \hline 
\cite{gaboardi2018psi, murtagh2018usable} & \ReOutreach & $\GeoUS$ & \SPtech & 20 & \SItaskCompletion & \checkmark & {\1}&{\1} & {\2} & {\2} & {\2} & {\2}  \\ \hline 
\cite{marcusland2020} &  \ReNetwork & $\GeoEurope$ & \SPtechDP & 22 & \SIsurvey, \SIinterview, \SItaskCompletion, $\EDwithinSubjects$ & \checkmark & {\2}&{\2} & {\0} & {\1} & {\0} & {\0}  \\ \hline 
\cite{alrige2013understanding} & \ReOutreach & $\GeoUS$ & \SPtech & 8 & \SIsurvey &  & {\0}&{\0} & {\0} & {\0} & {\0} & {\1}  \\ \hline 
\cite{nanayakkara2024measure} & \ReOutreach, \ReNetwork & $\GeoUS$ & \SPtech & 14 & $\SItaskCompletion, \SIinterview$  & \checkmark & {\1}&{\1} & {\2} & {\1} & {\1} & {\0}  \\ \hline 
\cite{sarathy2023don} & \ReNetwork & $\GeoUS$ & \SPtech & 19 & \SIsurvey, \SIinterview, \SItaskCompletion  &  & {\1}&{\1} & {\2} & {\0} & {\1} & {\2}  \\ \hline 
\hline
\multicolumn{13}{|l|}{
  \textbf{Recruitment key:}\;
  \begin{tabular}{r l r l}
      $\ReNetwork$:& Network &
      $\ReOutreach$:& Outreach 
  \end{tabular}}\\
  \hline
\multicolumn{13}{|l|}{
  \textbf{Geography key:}\;
  \begin{tabular}{r l l l}
      $\GeoUS$:& United States &
      $\GeoEurope$:& Europe 
  \end{tabular}}\\
  \hline
\multicolumn{13}{|l|}{
  \textbf{Sample Populations key:}\;
  \begin{tabular}{r l r l r l}
      $\SPtech$:& Technical users &
      $\SPtechDP$:& Technical users with DP expertise &
      $\SPend$:& End users 
  \end{tabular}}\\
  \hline
\multicolumn{13}{|l|}{
  \textbf{Study design \& instruments key:}\;
  \begin{tabular}{r l r l r l r l}
      $\SIsurvey$:& Survey &
      $\SIfocusGroup$:& Focus Group &
      $\SIinterview$:& Interview \\
      $\SItaskCompletion$:& Task-based Usability Test &
      $\EDbetweenSubjects$: & Between-subjects &
      $\EDwithinSubjects$:& Within-subjects
  \end{tabular}}\\

\hline
\end{tabular}
\caption{Methodologies used in studies of DP software tools.}

\label{tab:methodology_Studies_Investigating_Software_Tools}

\end{table*}

\begin{table*}
\newcommand{\RECRUITMENT}[1]{\begin{minipage}{9em}#1\end{minipage}}
\begin{tabular}{|c|c|c|p{3.2cm}|p{1.6cm}|c|c|l|c|c|c|c|c|}
\hline
  \multirow{2}{*}{\spheading{\textbf{Citation}}} &
  \multirow{2}{*}{\spheading{\textbf{Recruitment}}} &
  \multirow{2}{*}{\spheading{\textbf{Geo-location}}} &
  \multirow{2}{*}{\spheading{\textbf{Sample \\Populations}}} &
  \multirow{2}{*}{\spheading{\textbf{Sample Size}}} &
  \multirow{2}{*}{\spheading{\textbf{Study Methods\\ \& Instruments}}} &
  \multirow{2}{*}{\spheading{{\textbf{Educational Materials}}}} &
  \multicolumn{5}{|c|}{\textbf{Evaluation Methods}} \\ \cline{8-12}
  &&&&&&&
  \spheading{\textbf{Objective Understanding}} &
  \spheading{\textbf{Subjective Understanding}} &
  \spheading{\textbf{Satisfaction}} &
  \spheading{\textbf{Willingness to Share}} &
  \spheading{\textbf{Perception}}\\
\hline
\cite{franzen2022private} & $\ReAMT$
                         & $\GeoUS$
                         & \SPend
                         & 343
                         & $\SIsurvey$, $\EDbetweenSubjects$
                         & {}
                         & \filledcircle
                         & \filledcircle
                         & \opencircle & \opencircle & \halffilledcircle \\ \hline
\cite{cummings2021need} & $\ReAMT$
                        & $\GeoUS$
                        & \SPend 
                        & 1216
                        & $\SIsurvey$, $\EDbetweenSubjects$
                        &
                        & \opencircle
                        & \halffilledcircle
                        & \opencircle & \halffilledcircle & \halffilledcircle\\ \hline
\cite{10.1145/3576915.3623152} & $\ReProlific$
                        & $\GeoUS$
                        & \SPend 
                        & 22
                        & $\SIinterview$
                        & \checkmark
                        & \halffilledcircle
                        & \halffilledcircle
                        & \opencircle & \opencircle & \opencircle \\ \hline
\cite{9629406} & $\ReNetwork$
                        & $\GeoEurope$
                        & \SPtech
                        & 8
                        & $\SItaskCompletion$, $\SIfocusGroup$
                        & \checkmark
                        & \opencircle
                        & \opencircle
                        & \opencircle & \filledcircle & \opencircle \\ \hline
\cite{karegar2022exploring} & $\ReProlific$
                        & $\GeoEurope$
                        & \SPend
                        & 30
                        & $\SItaskCompletion$
                        & \checkmark
                        & \opencircle
                        & \filledcircle
                        & \opencircle & \filledcircle & \filledcircle \\ \hline
\cite{xiong2023exploring} & $\ReAMT$
                         & $\GeoUS$
                         & \SPtechDP, \SPend 
                         & 300
                         & $\SIsurvey$, $\EDbetweenSubjects$
                         & \checkmark
                         & \halffilledcircle 
                         & \filledcircle
                         & \opencircle & \halffilledcircle & \filledcircle \\ \hline
\cite{garrido2022lessons} & $\ReNetwork$ 
                         & $\GeoEurope$
                         & \SPtech 
                         & 24
                         & $\SIsurvey$
                         & \checkmark
                         & \opencircle
                         & \opencircle
                         & \filledcircle & \halffilledcircle & \opencircle \\ \hline
\cite{gaboardi2018psi} & $\ReNetwork$ 
                         & $\GeoUS$
                         & \SPtech
                         & 20
                         & $\SIsurvey$, $\SItaskCompletion$
                         & \checkmark
                         & \halffilledcircle
                         & \opencircle
                         & \opencircle & \halffilledcircle & \opencircle \\ \hline
\cite{kuhtreiber2022replication} & $\ReAMT$
                        & $\GeoEurope$
                        & \SPend
                        & 990 
                        & $\SIsurvey$, $\EDbetweenSubjects$
                        & \checkmark
                        & \halffilledcircle
                        & \halffilledcircle
                        & \opencircle & \halffilledcircle& \opencircle \\ \hline
\cite{wen2023influence} & $\ReProlific$
                        & $\GeoUS, \GeoEurope$
                         & \SPtechDP, \SPend
                         & 28
                         & $\SIsurvey$, $\EDbetweenSubjects$
                         & \checkmark 
                         & \halffilledcircle
                         & \halffilledcircle
                         & \halffilledcircle & \halffilledcircle &  \halffilledcircle \\ \hline
\multirow{4}{*}{\cite{xiong2020towards}}
    & \multirow{4}{*}{$\ReAMT$}
    & \multirow{4}{*}{$\GeoUS$}
    & \multirow{4}{*}{\SPend}
                        & 465
                        & $\SIsurvey$, $\EDbetweenSubjects$
                        & {}
                        & \halffilledcircle
                        & \halffilledcircle
                        & \opencircle & \halffilledcircle & \opencircle \\ \cline{5-8}
    & & & 
                        & 581
                        & $\SIsurvey$, $\EDbetweenSubjects$
                        & \checkmark
                        & \halffilledcircle
                        & \halffilledcircle
                        & \opencircle & \halffilledcircle & \opencircle \\ \cline{5-8}
    & & & 
                        & 468
                        & $\SIsurvey$, $\EDbetweenSubjects$
                        & \checkmark
                        & \halffilledcircle
                        & \halffilledcircle
                        & \opencircle & \halffilledcircle & \opencircle \\ \cline{5-8}
    & & & 
                        & 278
                        & $\SIsurvey$, $\EDbetweenSubjects$
                        & \checkmark
                        & \halffilledcircle
                        & \halffilledcircle
                        & \opencircle & \halffilledcircle & \opencircle \\ \cline{5-8} 
    & & & 
                        & 540
                        & $\SIsurvey$, $\EDbetweenSubjects$
                        & \checkmark
                        & \halffilledcircle
                        & \halffilledcircle
                        & \opencircle & \halffilledcircle & \opencircle \\ \hline
\cite{bullek2017towards} & $\ReAMT$
                        & $\GeoUS$
                        & \SPend
                        & 228
                        & $\SIsurvey$, $\EDwithinSubjects$, $\EDbetweenSubjects$
                        & \checkmark
                        & \opencircle
                        & \halffilledcircle
                        & \halffilledcircle & \halffilledcircle &  \\ \hline
\cite{smart2022understanding} & $\ReQualtrics$
                        & $\GeoUS$
                        & \SPend
                        & 300
                        & $\SIsurvey$, $\EDbetweenSubjects$
                        & \checkmark
                        & \filledcircle
                        & \filledcircle
                        & \filledcircle & \halffilledcircle & \opencircle \\ \hline
\cite{murtagh2018usable} & $\ReOutreach$ 
                        & $\GeoUS$
                        & \SPtech
                        & 28
                        & $\SItaskCompletion$
                        & 
                        & \halffilledcircle
                        & \halffilledcircle
                        & \halffilledcircle & \opencircle & \halffilledcircle \\ \hline
\cite{xiong2022using} & $\ReAMT$
                        & $\GeoUS$
                        & \SPtechDP, \SPend
                        & 300, 295 
                        & $\SIsurvey$, $\EDwithinSubjects$, $\EDbetweenSubjects$
                        & \checkmark
                        & \halffilledcircle
                        & \halffilledcircle
                        & \opencircle & \halffilledcircle & \opencircle \\ \hline
\cite{nanayakkara2023chances} & $\ReProlific$ 
                         & $\GeoUS$
                         & \SPend
                         & 963
                         & $\SIsurvey$, $\EDbetweenSubjects$ 
                         & \checkmark
                         & \halffilledcircle
                         & \halffilledcircle
                         & \halffilledcircle & \halffilledcircle & \opencircle \\ \hline
\multirow{2}{*}{\cite{valdez2019users}} 
                         & $\ReNetwork$ 
                         & $\GeoEurope$
                         & \SPend 
                         & 18
                         & $\SIfocusGroup$ 
                         & {}
                         & \opencircle
                         & \opencircle
                         & \opencircle & \halffilledcircle & \opencircle \\ \cline{2-8}
    & $\ReOutsource$ 
                         & $\GeoEurope$
                         & \SPend 
                         & 243 
                         & $\SIsurvey$, $\EDbetweenSubjects$
                         & {}
                         & \opencircle
                         & \opencircle 
                         & \halffilledcircle & \opencircle & \filledcircle \\ \hline
\cite{franzen2024communicating} & $\ReAMT$ 
                         & $\GeoUS$
                         & \SPend
                         & 308
                         & $\SIsurvey$, $\EDbetweenSubjects$ 
                         & \checkmark
                         & \filledcircle
                         & \opencircle
                         & \opencircle & \filledcircle & \opencircle \\ \hline
\cite{nanayakkara2022visualizing} & $\ReOutreach$ 
                         & $\GeoUS$
                         & \SPend
                         & 16
                         & $\SItaskCompletion$, $\EDwithinSubjects$, $\SIinterview$
                         & \checkmark
                         & \filledcircle
                         & \filledcircle
                         & \opencircle & \opencircle & \opencircle \\ \hline
\multicolumn{12}{|l|}{
  \textbf{Recruitment key:}\;
  \begin{tabular}{r l r l r l r l}
      $\ReAMT$:& Amazon Mechanical Turk &
      $\ReProlific$:& Prolific &
      $\ReQualtrics$:& Qualtrics \\
      $\ReNetwork$:& Network &
      $\ReOutreach$:& Outreach &
      $\ReOutsource$:& Contractor
  \end{tabular}}\\
  \hline
\multicolumn{12}{|l|}{
  \textbf{Geography key:}\;
  \begin{tabular}{r l r l r l r l}
      $\GeoUS$:& United States &
      $\GeoEurope$:& Europe 
  \end{tabular}}\\
  \hline
\multicolumn{12}{|l|}{
  \textbf{Sample Populations key:}\;
  \begin{tabular}{r l r l r l }
      $\SPtech$:& Technical users &
      $\SPtechDP$:& Technical users with DP expertise &
      $\SPend$:& End users 
  \end{tabular}}\\
  \hline
\multicolumn{12}{|l|}{
  \textbf{Study design \& instruments key:}\;
  \begin{tabular}{r l r l r l r l}
      $\SIsurvey$:& Survey &
      $\SIfocusGroup$:& Focus Group &
      $\SIinterview$:& Interview \\
      $\SItaskCompletion$:& Task-based Usability Test &
      $\EDbetweenSubjects$:& Between-subjects &
      $\EDwithinSubjects$:& Within-subjects &
  \end{tabular}}\\
  \hline

\end{tabular}
\caption{Methodologies used in studies communicating DP}
\label{tab:methodology_communicating_dp}
\end{table*}

\subsection{Recruitment and Sample}
\paragraph{Recruitment methods.} \label{sec:methodology:recruitment_method}
The "Recruitment" column in Tables~\ref{tab:methodology_Studies_Investigating_Software_Tools}
and~\ref{tab:methodology_communicating_dp} outlines the specific methods that these DP usability studies used to recruit study participants.
Studies evaluating DP software tools predominantly recruited participants through professional networks and direct outreach. Because these studies primarily targeted \textbf{technical users} of DP and did not require large samples, these tailored recruitment methods are suitable to reach potential participants with technical or data science skills required for study participation.
For example, Ngong et al.~\cite{ngong2024evaluating} leveraged email lists, Slack channels, and professional networks. PSI studies \cite{lee2018salience, murtagh2018usable} used social science listservs and student job
boards at local universities.

In comparison, studies focusing on communicating DP employed a variety of recruitment methods. Some study also leveraged professional networks \cite{9629406, garrido2022lessons, gaboardi2018psi} and outreach efforts \cite{murtagh2018usable} to reach \textbf{technical users}. However, most of these studies \cite{franzen2022private, nanayakkara2023chances, smart2022understanding} resorted to online crowdwork platforms (e.g., Amazon Mechanical Turks, Prolific, Qualtrics) because they sought potential \textbf{end users} of DP. These platforms enabled researchers to quickly recruit large samples from the general public with the ability to control certain demographic variables. 

In summary, most DP usability studies chose appropriate recruitment methods to reach the their targeted populations (see \ref{sec:methodology:sample_expertise}) and the sample sizes suitable for their study designs (see \ref{sec:methodology:sample_size}). However, current recruitment methods in DP usability studies lack width and diversity. Specifically, researchers largely relied on their own professional networks to recruit \textbf{technical users} and on crowdwork platforms to recruit \textbf{end users}. Despite the convenience, inherent sampling biases from small professional networks and participants on crowd-sourcing platforms are unavoidable. In our opinion, diversifying recruitment strategies in future DP usability studies will increase the overall validity and reliability of research in this area.

\paragraph{Geo-location.}\label{sec:methodology:geo-location} 
The "Geo-location" column revealed that all DP usability studies in Tables~\ref{tab:methodology_Studies_Investigating_Software_Tools}
and~\ref{tab:methodology_communicating_dp} were conducted by researchers in the US and the EU with samples from developed Western countries. We believe reaching participants across countries and cultures will improve the external validity of future DP usability studies, particularly for studies targeting end users.

\paragraph{Sample Populations.}
\label{sec:methodology:sample_expertise}
The "Sample Populations" column in Tables~\ref{tab:methodology_Studies_Investigating_Software_Tools}
and~\ref{tab:methodology_communicating_dp} shows that DP usability studies in this SoK recruited end users, technical users, and technical users with DP expertise.
Generally, the research questions of a specific study determined the target sample populations. Studies on DP communication primarily recruited end users from crowdwork  channels  like Amazon Mechanical Turk, Prolific, Qualtrics etc. because many seek to examine the effectiveness of DP explanation and user understanding where end users with diverse background are appropriate.
In addition, some DP communication studies \cite{wen2023influence, xiong2022using, xiong2023exploring} also involves technical users because the authors wanted to get perspective from users from various technical background for better understanding and comparison with end users. Also, some communication studies only used technical users because they require DP knowledge to interpret certain study materials~\cite{gaboardi2018psi, garrido2022lessons, murtagh2018usable}.

In contrast, studies examining DP tools often require participants with certain technical backgrounds, such as \cite{gaboardi2018psi} to be able to use DP tools. Specifically, we observe that some studies \cite{ marcusland2020, ngong2024evaluating}  further differentiate technical users by their prior DP expertise  This is necessary because those technical users integrate software tools in their product pipeline to add DP to their existing system.

\paragraph{Sample size.}
\label{sec:methodology:sample_size}
The sample sizes in these DP usability studies vary greatly but generally align with their chosen study methods (detailed in \ref{sec:methodology:study_design}).
Studies that involved in-depth qualitative methods (e.g. focus groups, interviews) or time-consuming task-based evaluation had small sample sizes of no more than 30 participants. These include all studies evaluating DP software tools (Tables~\ref{tab:methodology_Studies_Investigating_Software_Tools}) and a handful of studies that focused on communicating DP \cite{9629406, 10.1145/3576915.3623152, karegar2022exploring, gaboardi2018psi, garrido2022lessons, murtagh2018usable, nanayakkara2022visualizing, valdez2019users, wen2023influence}.
Studies that employed the web-based survey method \cite{franzen2022private, cummings2021need, xiong2023exploring, kuhtreiber2022replication, xiong2020towards, bullek2017towards, smart2022understanding, xiong2022using, nanayakkara2023chances,valdez2019users,franzen2024communicating} had relatively large sample sizes, from a few hundred to more than a thousand participants. Most of these studies focused on communicating DP to end users and leveraged crowdwork platforms for recruitment. 

Studies targeting technical users ~\cite{9629406,alrige2013understanding,nanayakkara2024measure, sarathy2023don, xiong2023exploring}, including individuals with data science background, programming skills or prior DP knowledge, tended to have small samples. This is partially  due to the difficulty to recruit participants with technical expertise. 
Overall, mostly of the DP usability studied reviewed in this SoK had adequate sample sizes compared to published work in usable security and privacy research area utilizing similar study methods.

\subsection{Study Methods \& Instruments} \label{sec:methodology:study_design}
DP usability studies reviewed in this SoK employed different research methods and various instruments to gather quantitative and qualitative data. Below, we describe how these methods and instruments were used in DP usability studies. 

\paragraph{Survey ($\SIsurvey$).} 
Surveys are structured questionnaires to collect quantitative and qualitative data relative large samples. This method is suitable to gauge user perceptions, comprehension, and decision-making, or perform web-based experiments. For instance, \cite{franzen2022private, cummings2021need} utilized surveys to evaluate DP communication effectiveness and user expectations. Survey questionnaire can also be a supplemental instrument for other methods. For example, Ngong et al.\cite{ngong2024evaluating} used surveys in task-based usability test to gauge pre-task and post-task data from participants. 

\paragraph{Interview ($\SIinterview$).}  Interviews enable in-depth exploration of participants' experiences, reasoning, and feedback, making them valuable for qualitative insights. For example Nanayakkara et al. \cite{nanayakkara2023chances} and Xiong et al. \cite{xiong2022using} both used this approach to compare more versus less visually descriptive DP representations.

\paragraph{Focus Group ($\SIfocusGroup$).} 
Focus groups gather qualitative insights into collective perceptions and attitudes from a group of participants. For example, \cite{9629406, valdez2019users} used focus groups to explore privacy risks and trade-offs, informing better communication strategies.


\paragraph{Task-based Usability Test ($\SItaskCompletion$).} 
Task completion exercises are a research instrument to assess a tool's usability by observing users complete key tasks using the tool.
Collected data include success rates, efficiency, or qualitative observations of user behavior. For example, \cite{ngong2024evaluating, 9629406, marcusland2020} used this method to evaluate user performance and decision-making with differential privacy tools.

\paragraph{Experiment ($\EDwithinSubjects$ and $\EDbetweenSubjects$).} 
Experiments are a research method to compare the results of at least two groups of users in different controlled conditions. The study design of experiments can be within-subject or between-subject.
Within-subject design (denoted by $\EDwithinSubjects$) involves each participant in multiple conditions, allowing for comparative analysis of different tool configurations with relative small samples. For example, Bullek et al. \cite{bullek2017towards} employed this approach to compare user preferences across multiple settings.
Between-subject design (denoted by $\EDbetweenSubjects$) ensures each participant is only exposed to one condition, which is beneficial for isolating the effect of the specific condition. This design is well-suited for studies utilizing crowdwork platforms, where larger, more diverse samples can be recruited, to assign each participant into a unique condition.
Most DP usability studies utilizing experiments adopted a between-subjects design (e.g.~\cite{bullek2017towards, cummings2021need, franzen2024communicating, franzen2022private, kuhtreiber2022replication, nanayakkara2023chances, ngong2024evaluating, smart2022understanding, valdez2019users, wen2023influence, xiong2020towards, xiong2022using, xiong2023exploring}).

\paragraph{Educational materials.} Some DP usability studies utilized a special type of research instrument to educate participants with basic DP knowledge before study procedures, as shown in the "Educational Materials" column of Tables~\ref{tab:methodology_Studies_Investigating_Software_Tools} and~\ref{tab:methodology_communicating_dp}.
For example, Gaboardi et al. \cite{gaboardi2018psi} and Murtagh et al. \cite{murtagh2018usable} incorporate educational resources like handouts, tutorials, or short videos to establish a ``common frame of reference'' among participants. 
DP educational materials help establish a consistent baseline of knowledge among participants, when such knowledge is needed for the study procedure.

Many studies in this SoK employed various methods in pursuit of variations of the research question: Can different groups of users understand and properly apply differential privacy given a range of realistic scenarios? Table 3 lists best practices for Experiment Design \& Instrument Type. Next-generation DP usability studies will use the same basic set of methods and instruments, but will also benefit from greater adoption of DP, which in turn should provoke focused research questions that help solve critical DP issues.

\subsection{Evaluation Metrics} \label{sec:method_evalaution}
The DP usability studies reviewed in this SoK used various evaluation metrics to assess the usability of DP tools and the effectiveness of DP communication. 

\iparagraph{Shared metrics.}
Both studies on DP tools and DP communication measured the following aspects regarding DP's usability.

\paragraph{Objective understanding.}
This involves quantitative or qualitative measurements of participants' knowledge about what DP terms mean, their ability to reason about DP, or their ability to solve DP tasks. For example, Ngong et al.~\cite{ngong2024evaluating} assessed objective understanding by evaluating participants' abilities to accurately interpret DP parameters and apply DP concepts in practical scenarios.
This is typically measured through pre- and post-study quizzes, task-based assessments, and scenario-based questions, which help accurately gauge participants' understanding with standard questions.

\paragraph{Subjective understanding.} 
This refers to how participants conceptualize DP, including their mental models, personal interpretations, concerns, and ethical intuitions about DP~\cite{cummings2021need}. Researchers assess it using qualitative methods such as interviews, focus groups, and self-reported surveys~\cite{franzen2022private, nanayakkara2023chances}. Nanayakkara et al. ~\cite{nanayakkara2023chances} echo methods applied previously in Franzen et al.~\cite{franzen2022private}, where they measured self-assessments of confidence for DP understanding by Likert scales and analyzed statistically. Nanayakkara et al. ~\cite{nanayakkara2023chances} took the extra step to have subjects describe DP privacy protection in their own words. While subjective understanding provides insights into how individuals internalize and process DP concepts, personal biases and varying levels of technical literacy may influence responses, requiring careful interpretation.

\paragraph{Satisfaction.} 
This measures how satisfied participants are that differential privacy provides adequate protection for personal information or that a fair balance between privacy and utility has been struck. This is measured through post-study surveys and interviews to gather participants' satisfaction ratings and qualitative feedback. Satisfaction surveys are effective for gauging overall user sentiment, though they can be subjective. Cummings et al. \cite{cummings2021need} equated satisfaction with subjects' met (or unmet) expectation for privacy protection based on task outcome. Ngong et al. \cite{ngong2024evaluating} used common recommendation comparison metrics (Net Promoter Score, System Usability Score) to compare satisfaction levels for different DP tools.

\iparagraph{Metrics for DP tools.}
Many studies on DP tools used classic software usability metrics to evaluate DP tools, as detailed below.

\paragraph{Task success rate.} 
This measures how often participants complete tasks using DP tools. This metric is crucial for understanding the practical usability of the tools. It is typically measured through direct observation and logging of task completion rates during usability testing sessions. Task success rate is a straightforward and reliable metric for assessing practical usability. Quite a few papers assessed study participants' understanding of DP through task completion \cite{9629406, karegar2022exploring, gaboardi2018psi, murtagh2018usable, nanayakkara2022visualizing, ngong2024evaluating, nanayakkara2024measure, sarathy2023don}. A subset of these studies used talk-aloud methods to capture participants' thought process and decision-making \cite{ngong2024evaluating, nanayakkara2024measure, nanayakkara2022visualizing, sarathy2023don}.

\paragraph{Time on task.} 
This quantifies how long participants take to complete tasks
and reflects "efficiency" for tool-based problem solving according to Nielsen's "5 Components of Usability" \cite{nielsen1994measuring}. Efficiency encapsulates the speed and ease of the workflow leading to task completion. Systems that are inefficient in comparison can help identify usability issues or learning curves associated with the tools. Studies by Ngong et al. \cite{ngong2024evaluating} and Murtagh et al. \cite{murtagh2018usable} included time on task as a key metric to evaluate the efficiency of the tools being tested. This is measured by timing participants during task completion exercises. This metric effectively highlights efficiency and potential areas where the user experience can be improved.

\paragraph{Error rate.} 
Error rate measures the frequency of errors made by participants while using the tools. Pilot studies help to scope possible errors in advance of formal investigation, especially since error rates depend on having clear definitions for what does and does not constitute a task- or tool-related error. Several studies~\cite{alrige2013understanding, marcusland2020, ngong2024evaluating} used error rate to gauge how often users make mistakes and the types of errors encountered.

\iparagraph{Metrics for DP communication.}
Some DP communication studies also measured other subjective constructs of end users.

\paragraph{Willingness to share.} 
This evaluates how willing participants are to contribute their own private data in scenarios where differential privacy (DP) techniques are applied. This is often measured through surveys and interviews, where participants are asked about their comfort levels and willingness to share personal data under various conditions. Several studies have examined participants' willingness to share data after being informed about DP implementations, finding that clear explanations and illustrations significantly influenced their willingness to share data \cite{xiong2022using, xiong2023exploring, xiong2020towards, wen2023influence,bullek2017towards}.

\paragraph{Perception.}     
This evaluates participants' trust, confidence, and attitudes toward DP implementations and the organizations deploying them. It examines how participants judge the appropriateness, fairness, and effectiveness of real-world DP deployments. It assesses how people feel about DP applications and their confidence in the privacy guarantees they provide. Hands on studies with interactive DP tools like DP Creator \cite{sarathy2023don}, ViP \cite{nanayakkara2022visualizing}, PSI \cite{gaboardi2018psi, murtagh2018usable} and DPP tool \cite{9629406} compelled users to determine for themselves appropriate DP parameters like privacy budget and epsilon in order to complete a simulated task. Each study followed up the exercises with interviews or focus groups to fully understand users' perception of DP parameter appropriateness.

\section{Results}
\label{sec:results_usability}

\subsection{DP Software Tools}
\label{sec:software-tools}
This section corresponds with Table~\ref{tab:software_tools}, which summarizes the findings from various studies that investigated different software tools designed for Differential Privacy (DP). 
The table highlights key aspects such as the types of tools developed, the target users' expertise and role, the support for privacy budgeting, utility analysis, level of automation, flexibility, correctness checking, and deployment model. These elements are crucial in evaluating how well these tools facilitate the implementation and understanding of DP across diverse user groups.

\begin{table}

\newcommand{\Visual}{V}
\newcommand{\API}{A}
\newcommand{\Hypothetical}{H}
\begin{tabular}
{|p{14pt}|c|p{30pt}|p{35pt}|c|c|c|c|c|p{15pt}|}
\hline
  \spheading{\textbf{Citation}} &
  \spheading{\textbf{Type of tool}} &
  \spheading{\textbf{Target expertise}} &
  \spheading{\textbf{Target Role}} &
  \spheading[80pt]{\textbf{Parameter-setting}} &
  \spheading{\textbf{Utility analysis}} &
  \spheading{\textbf{Automation}} &
  \spheading{\textbf{Flexibility}} &
  \spheading[90pt]{\textbf{Correctness checking}} &
  \spheading{\textbf{Deployment model}} \\ \hline
  
   \cite{9629406}
   & \Visual
   & \SPtech, \SPtechDP
   & Curators
   & {\2}
   & {\1}
   & {\0}
   & {\2}
   & {\0}
   & LDP

   \\ \hline

   \cite{ngong2024evaluating}
   & \API
   & \SPtech, \SPtechDP
   & Curators, analysts
   & {\2}
   & {\0}
   & {\0}
   & {\2}
   & {\0}
   & CDP
   \\ \hline

      \cite{garrido2022lessons}
   & \Hypothetical
   & \SPtechDP
   & Analysts
   & {\2}
   & {\0}
   & {\0}
   & {\2}
   & {\2}
   & LDP, CDP
   \\ \hline

      \cite{gaboardi2018psi, murtagh2018usable}
   & \Visual
   & \SPtech, \SPtechDP
   & Owners, curators, analysts
   & {\2}
   & {\2}
   & {\0}
   & {\0}
   & {\0}
   & CDP
   \\ \hline

    \cite{lobner2023user}
   & \Hypothetical
   & \SPtech
   & Curators, analysts
   & {\1}
   & {\0}
   & {\1}
   & {\1}
   & {\0}
   & LDP, CDP
   \\ \hline

    \cite{nanayakkara2024measure}
   & \Visual
   & \SPtech, \SPtechDP
   & Curators, analysts
   & {\2}
   & {\2}
   & {\1}
   & {\1}
   & {\1}
   & CDP
   \\ \hline

    \cite{sarathy2023don}
   & \Visual
   & \SPtech, \SPtechDP
   & Curators, analysts
   & {\2}
   & {\1}
   & {\2}
   & {\0}
   & {\2}
   & CDP
   \\ \hline

\multicolumn{10}{|l|}{\textbf{Type of tool key:} \Visual: Visual, \API: API-based, \Hypothetical: Hypothetical} \\ \hline
\multicolumn{10}{|l|}{\textbf{Target expertise key:} \phantom{hello}$\SPtech$: Technical users} \\
\multicolumn{10}{|l|}{
\phantom{hi}$\SPtechDP$: Technical users with DP expertise 
\phantom{hi}$\SPend$: End users } \\ \hline

\end{tabular}

\caption{Key aspects examined in studies of DP software tools}
\label{tab:software_tools}
\vspace*{-20pt}
\end{table}

\iparagraph{Tool Design.}
\label{type-of-tool}
\paragraph{API tools.} API tools assist practitioners in implementing DP solutions by writing code in a mainstream programming language. These tools are designed to be used to prepare DP data releases or implement new DP systems, without needing to implement DP mechanisms from scratch.
%

\paragraph{Visual tools.}
Visual tools for DP are designed to provide a user-friendly interface for preparing a DP data release without writing any code. These tools often include visual aids and interactive elements that guide users through the process of setting privacy parameters and analyzing data, making them particularly valuable for end users and technical users without DP expertise~\cite{gaboardi2018psi,9629406, sarathy2023don}.

\paragraph{Hypothetical tools.}
Some tools have been proposed, but not implemented. Some of these hypothetical tools align with the current implementation of API or visual tools~\cite{garrido2022lessons}. Others are entirely new~\cite{murtagh2018usable, lobner2023user}, and results of users studies on mockups of these tools can provide valuable guidance for tool design.

\paragraph{Target role: who uses the tool?}
DP systems and DP data releases involve collaboration between people in different roles, and different tools are designed to help different roles. Tools may be designed to help the \emph{data analyst}, who designs and implements the pipeline that takes sensitive data as input and produces DP results as output; the \emph{data subject} (sometimes called \emph{data owner}), who submits the sensitive data for analysis in the first place; the \emph{data curator} (or \emph{data steward}), who accept data submissions and maintain the database of sensitive data.
Most existing tools are designed to help the data analyst to implement software pipelines that successfully enforce DP with the desired privacy parameters and level of utility~\cite{lobner2023user, murtagh2018usable, gaboardi2018psi, garrido2022lessons, ngong2024evaluating}. A subset of tools are also designed to help the data curator~\cite{murtagh2018usable, sarathy2023don}.

\paragraph{Deployment model.}
Current tools assume the central model of DP, in which the sensitive data is collected on a central server by a trusted data curator. Tool support for other deployment models, including the local and shuffle models, is currently lacking.

\iparagraph{Target Expertise.} Many tools claim to improve usability for technical users without DP expertise. Studies of these tools group participants according to DP expertise.

\paragraph{Tech:} This group includes technical users \emph{without} DP expertise. These users benefit the most from visual tools that provide a user-friendly interface, require minimal DP knowledge and guide users towards the right usage of DP. Tools like PSI~\cite{gaboardi2018psi} and DP Creator~\cite{sarathy2023don} are particularly valuable for this group as they simplify the process of setting privacy parameters through editable forms and visual aids, reducing the need for DP expertise.

\paragraph{TechDP:}
This group includes technical users \emph{with} DP expertise. These users are the typical target audience for API-based tools that offer extensive customization and flexibility. Such tools are designed to integrate with various programming languages and data processing frameworks.


\iparagraph{Support for Setting Parameters.}
%
\noindent Every API tool~\cite{ngong2024evaluating} and most visual tools~\cite{gaboardi2018psi, sarathy2023don} we examined provided a method for setting the privacy budget ($\varepsilon$ or corresponding parameter(s)).

Visual tools like PSI~\cite{gaboardi2018psi} and DP Creator~\cite{sarathy2023don} provide user-friendly interfaces for setting privacy budgets. These tools offer default options and allow users to allocate the total privacy budget across different queries through editable forms. PSI, for example, lets users reserve a portion of the budget for specific queries, with the rest divided among data attributes, while DP Creator allows for updating privacy parameters for different dataset entities.

API-based tools offer more flexibility for technical users with DP expertise users to set and adjust the privacy budget programmatically. This flexibility, while powerful, increases complexity and requires a good understanding of both DP concepts and the API itself to avoid misconfigurations~\cite{ngong2024evaluating}. 
In all cases, users need to understand the privacy parameters, either through their own knowledge of DP or by consulting an expert---a challenging task, as indicated by studies on tools~\cite{sarathy2023don, ngong2024evaluating} and the communication studies described in Section~\ref{sec:Communicating_DP}.

Some hypothetical tools aim to further improve support for budgeting by automating the process and providing clearer guidance to users. For example, the hypothetical tools discussed by Murtagh et al.~\cite{murtagh2018usable} do not recommend specific privacy parameters but instead focus on providing mechanisms to estimate the risk of data sharing.

Another important aspect of support for budgeting is the setting of clipping parameters, which define the range of allowable data values. Both visual~\cite{gaboardi2018psi, sarathy2023don} and API tools~\cite{ngong2024evaluating} allow users to set these bounds, and they often include warnings to highlight the importance of appropriate clipping to maintain both privacy and utility. Inappropriate clipping parameters can lead to poor data utility, especially if they affect a significant portion of the data.

\paragraph{Support for utility analysis.}
Utility analysis should tell us if the DP release provides sufficient accuracy to enable the desired downstream analyses. Some tools provide accuracy metrics to help the target user understand the utility of the release and navigate the privacy/utility tradeoff~\cite{gaboardi2018psi, sarathy2023don}.
Support for this kind of analysis is important, since analysts report concern that the DP release will not be sufficiently accurate for the desired uses~\cite{garrido2022lessons}. 

PSI~\cite{gaboardi2018psi} calculates mean absolute error for every selected column on the output as utility metric, and allows adjusting the confidence level. DP Creator~\cite{sarathy2023don} provides a similar accuracy report, but is less transparent about how error is calculated.

\paragraph{Level of automation and flexibility.}
Level of automation for DP software tools indicate the tradeoff between how many parameter values have to be decided by user and how many the system will determine automatically. Both visual and API tools require the analyst to provide values for most parameters~\cite{gaboardi2018psi, sarathy2023don, ngong2024evaluating} (e.g. lower and upper bounds for clipping) and to select the precise mechanism to be used. In both contexts, analysts struggle with setting these parameters~\cite{ngong2024evaluating, sarathy2023don}, suggesting that increased automation would be helpful for users.

Level of flexibility refers how much control the user has in determining how DP will be achieved. Existing visual tools are less flexible: they provide wizard-like interfaces that support a small, fixed set of analyses~\cite{gaboardi2018psi, sarathy2023don}. API tools differ in their levels of flexibility~\cite{ngong2024evaluating}: some provide access to low-level mechanisms and enable users to build new mechanisms on top (e.g. OpenDP and DiffPrivLib), while others aim to provide a higher-level API that requires less expertise (e.g. Tumult Analytics). Existing tools (both visual and API-based) tend to require data to be stored in a particular format or underlying data store, limiting integration with existing data processing infrastructure.

\paragraph{Level of correctness checking.}
In the context of Differential Privacy (DP) tools, correctness checking refers to the tool's ability to ensure that the deployed system is free of privacy bugs. Interactive tools like DP Creator~\cite{sarathy2023don} prevent bugs by limiting available functionality to correct operations. In flexible API-based tools, ensuring correctness is more difficult. Ngong et al.~\cite{ngong2024evaluating} highlight that even experienced practitioners can inadvertently implement DP mechanisms incorrectly when using API-based DP tools. Despite their confidence, these users often made mistakes that resulted in insecure or flawed DP implementations. This finding underscores the necessity of built-in correctness-checking tools to ensure the accuracy and security of DP implementations.

\begin{table}
\begin{tabular}{|l|l|l|l|l|l|}
\hline
\multirow{3}{*}{\spheading{\textbf{Citation}}} &
  \multicolumn{2}{c|}{\textbf{Communicating}} &
  \multicolumn{3}{c|}{\textbf{Communicating}} \\ 
  & \multicolumn{2}{c|}{\textbf{about}} &
  \multicolumn{3}{c|}{\textbf{using}} \\ \cline{2-6} 
   &
\spheading{Epsilon or other parameters} &
  \spheading{Deployment\\model\\(CDP vs LDP)} &
  \spheading{Descriptions (text)} &
  \spheading{Visualizations} &
  \spheading{Pictures\\\& diagrams} \\ \hline
  
  \cite{franzen2022private} & \halffilledcircle & \opencircle & \filledcircle & \opencircle & \opencircle \\ \hline
  \cite{cummings2021need} & \halffilledcircle & \opencircle & \filledcircle & \opencircle & \opencircle \\ \hline
  \cite{10.1145/3576915.3623152} & \opencircle & \opencircle & \halffilledcircle & \halffilledcircle & \halffilledcircle \\ \hline
  \cite{karegar2022exploring} & \halffilledcircle & \filledcircle & \opencircle & \filledcircle & \opencircle \\ \hline
  \cite{xiong2023exploring} & \halffilledcircle & \filledcircle & \halffilledcircle & \opencircle & \filledcircle \\ \hline
  \cite{garrido2022lessons} & \opencircle & \halffilledcircle & \halffilledcircle & \opencircle & \opencircle \\ \hline
  \cite{steil2019privacy} & \opencircle & \opencircle & \halffilledcircle & \opencircle & \opencircle \\ \hline
  \cite{kuhtreiber2022replication} & \opencircle & \filledcircle & \filledcircle & \opencircle& \opencircle \\ \hline
  \cite{wen2023influence} & \filledcircle & \opencircle & \halffilledcircle & \opencircle &  \filledcircle \\ \hline
  \cite{xiong2020towards} & \filledcircle & \filledcircle & \filledcircle & \opencircle & \opencircle \\ \hline
  \cite{bullek2017towards} & \opencircle & \halffilledcircle & \halffilledcircle & \filledcircle & \opencircle \\ \hline
  \cite{smart2022understanding} & \filledcircle & \filledcircle & \filledcircle & \opencircle & \opencircle \\ \hline
  \cite{nanayakkara2023chances} & \filledcircle & \opencircle & \filledcircle & \filledcircle & \filledcircle \\ \hline
  \cite{valdez2019users} & \opencircle & \opencircle & \halffilledcircle & \opencircle & \filledcircle \\ \hline
  \cite{franzen2024communicating} & \filledcircle & \opencircle & \filledcircle & \filledcircle & \halffilledcircle \\ \hline
  \cite{nanayakkara2022visualizing} & \filledcircle & \opencircle & \opencircle & \filledcircle & \opencircle \\ \hline
\end{tabular}
\caption{Key aspects examined in studies communicating DP}
\label{tab:communicating-about}
\vspace*{-20pt}
\end{table}

\subsection{Communicating DP}
\label{sec:Communicating_DP}

\iparagraph{Communicating Using.}
Effective communication in differential privacy utilizes descriptions, visual aids - visualizations, and pictures/diagrams. Each method caters to different aspects of comprehension and informational needs.

\paragraph{Descriptions (Text).} Textual descriptions serve as the foundation for explaining DP concepts. They are most effective when they simplify complex ideas and relate them to everyday situations. Narratives that contextualize DP settings in familiar scenarios can significantly improve comprehension, as evidenced by the textual strategies recommended by Cummings et al. \cite{cummings2021need}. Their study highlights the importance of using simple, relatable
language to explain differential privacy, making it accessible to end users.
Franzen et al. \cite{franzen2022private} emphasize using risk communication formats to articulate the implications of DP parameters, such as epsilon ($\varepsilon$), clearly and directly. This approach helps users understand the trade-offs between privacy and utility, enhancing their overall comprehension. Xiong et al. \cite{xiong2023exploring} also support the use of clear textual explanations to describe DP models, noting that simplifying complex mathematical principles significantly aids non-technical audiences in understanding. Also, the work of Garrido et al. \cite{garrido2022lessons} and Kühtreiber et al. \cite{kuhtreiber2022replication} highlights the importance of detailed textual descriptions in improving users' comprehension and data-sharing decisions. These studies suggest that clear, concise descriptions can effectively communicate the nuances of DP, thereby increasing users' willingness to share their data.

Despite these advantages, textual descriptions alone may not always suffice to convey the complexities and nuances of differential privacy. 
 Additionally, the current descriptions of DP are often insufficient to help users make informed decisions, lacking consistency and standardization \cite{cummings2021need}. This emphasizes the need for new, standardized descriptions to improve user understanding and trust in DP implementations.

\iparagraph{Visualizations.}
Across studies, visualization has emerged as a natural strategy for communicating with end users.
Our review suggests that visualizations—especially icon arrays, metaphoric illustrations, and interactive sliders—are more effective than text alone in communicating DP concepts.

\paragraph{Visualizations of privacy.}
Franzen et al.~\cite{franzen2024communicating} evaluated five interface designs to support data donors in making privacy decisions in a between-subjects online experiment with 378 end users. The study measured \textit{concern-decision consistency}—the alignment between participants’ stated privacy concerns and their sharing choices—as the primary outcome. Results showed that interfaces incorporating visual risk representations led to more consistent and informed decision-making than text-only designs. 

Nanayakkara et al.~\cite{nanayakkara2023chances} developed and compared three explanation formats for communicating the privacy implications of the $\epsilon$ parameter in DP: a textual odds description (ODDS-TEXT), a visual odds representation using frequency-framed icon arrays (ODDS-VIS), and concrete output examples (SAMPLE REPORTS). The study was conducted as a randomized vignette-based online experiment with 963 end users. 
The evaluation measured objective risk comprehension, subjective privacy understanding, and willingness to share data. Results showed that ODDS-VIS  improved both objective and subjective understanding compared to the other formats. 

Wen et al.~\cite{wen2023influence} conducted a design-space exploration to evaluate how different explanation formats affect user understanding of Local Differential Privacy (LDP). They tested three formats: (1) textual descriptions, (2) data tables illustrating algorithmic behavior, and (3) explanatory illustrations, such as lottery-ball drawings designed to visualize probabilistic noise injection.
The study involved a large-scale randomized online survey with 228 end users.
Findings showed that participants in the illustration condition achieved significantly higher comprehension scores compared to those in the text or table conditions. The metaphor-driven visuals—particularly the lottery analogy—helped users form more accurate mental models of local noise mechanisms and privacy protection. 

Bullek et al.~\cite{bullek2017towards} investigated whether visualizing the obfuscation process in the Randomized Response technique (RR)—a mechanism consistent with Local Differential Privacy—could enhance users' trust and understanding of privacy guarantees. They developed a visual explanation illustrating how responses are randomly flipped to ensure privacy.
The study involved 228 participants. Outcome measures included self-reported trust, comfort with the mechanism, comprehension of how RR works, and willingness to share data.
Results showed that participants exposed to the visual explanation reported higher trust and greater comfort in using RR. However, some participants made riskier data-sharing choices despite grasping the mechanism. 

\paragraph{Visualizations of uncertainty for navigating the privacy-utility tradeoff.}
Nanayakkara et al.~\cite{nanayakkara2022visualizing} introduced the ViP (Visualizing Privacy) tool to support data curators in configuring $\epsilon$ for DP data releases. The tool featured interactive sliders and real-time visual feedback that dynamically illustrated how varying $\epsilon$ values affected both accuracy and disclosure risk, including changes in confidence interval width and the probability of extreme (noisy) outputs.
The tool was evaluated through a within-subjects study involving 16 clinical research practitioners with statistical training but limited familiarity with DP. Participants completed a series of $\epsilon$-setting tasks using both ViP and a spreadsheet-based control interface that lacked visualization. The study assessed participants’ accuracy in interpreting probabilistic outcomes, and  their ability to select $\epsilon$ values aligned with specified privacy or accuracy goals.
Results showed that ViP significantly improved performance on both measures. Participants using the tool demonstrated better understanding of the privacy-utility trade-off and made better parameter choices. 

Calero Valdez and Ziefle~\cite{valdez2019users} used a conjoint analysis (N=521) to study user preferences around data sharing in health recommender systems. Scenarios varied in privacy level (including DP), data type (mental vs. physical health), and purpose (scientific vs. commercial). While not focused on visualization per se, the study relied on interface sliders and privacy parameter displays. Participants favored stronger privacy guarantees and demonstrated contextual sensitivity, showing that visual presentation of privacy settings influences preferences even without explicit DP education.

\paragraph{Diagrams for describing deployment models.}
Karegar et al.~\cite{karegar2022exploring} developed a set of pictorial metaphors to explain the privacy mechanisms of Central and Local DP to technical users without DP expertise in the context of eHealth data sharing. These visuals used imagery such as blurred portraits and obscured shadows to convey concepts like data perturbation, anonymity, and the privacy-utility trade-off.
The evaluation was conducted through 30 semi-structured online interviews. Each participant was shown one or more metaphors tailored to specific DP models and asked to describe their understanding of how the mechanism protected their data. 
Results indicated that the metaphors effectively conveyed general ideas such as the presence of noise and the loss of identifiability. However, several participants developed overgeneralized or inaccurate mental models—such as assuming complete anonymity—particularly when visual cues were not anchored to real-world scenarios. 

Xiong et al.~\cite{xiong2023exploring} designed explanative illustrations to compare the three major DP models: Central DP, Local DP, and Shuffler DP. Each illustration included a textual explanation and a data flow diagram that visually conveyed the model’s trust assumptions, noise injection points, and privacy-utility trade-offs. The diagrams were designed using dual coding theory to reinforce learning through verbal and pictorial cues. A utility heatmap was also included to visualize the impact of DP mechanisms on data accuracy at the aggregate level.
The evaluation involved an online experiment with 300 end users recruited from Amazon Mechanical Turk. Participants were assessed on comprehension of each DP model’s mechanism, perceived privacy and utility, and willingness to share data in public-good and commercial-interest scenarios. 
Findings showed that participants exposed to the visual explanations achieved significantly higher comprehension of model structure and implications, particularly for Shuffler DP. The visualizations also improved perceived privacy protection and influenced data-sharing preferences. The study concluded that explanative illustrations can improve understanding of complex DP models and should be paired with textual scaffolding to support informed user decisions.

\iparagraph{Communicating About.}
Effectively communicating complex Differential Privacy (DP) concepts is crucial for fostering understanding and trust among users. This section discusses the best approaches to articulate key DP parameters and the differences between common deployment models such as Centralized Differential Privacy (CDP) and Local Differential Privacy (LDP).

Clarifying parameters like epsilon is essential for explaining DP's privacy guarantees. Table \ref{tab:communicating-about} summarizes various methods—such as text descriptions and visualizations—used in the studies we reviewed to communicate these concepts effectively.
Understanding the differences between CDP and LDP is vital for informed decision-making. Table \ref{tab:communicating-about} also highlights the use of text and visuals to enhance users' comprehension of these models.


\paragraph{Epsilon and other parameters.}
"Epsilon" ($\varepsilon$) is the primary parameter in differential privacy (DP) that quantifies the level of privacy guarantee, encapsulating the trade-off between privacy and data utility. 
Our literature review reveals that effectively communicating privacy parameters hinges on simplification and contextualization. Studies employ a variety of methods, with textual explanations or descriptions being the most prevalent \cite{cummings2021need, franzen2022private, xiong2020towards}. These studies such as those by Cummings et al.~\cite{cummings2021need}
and Franzen et al.~\cite{franzen2022private} demonstrate that presenting complex mathematical concepts in simpler terms significantly enhances comprehension among non-technical audiences. Franzen et al.~\cite{franzen2022private} advocate for the use of risk communication formats from the medical field to articulate $\varepsilon$'s implications clearly and directly.

Previous studies have been inconclusive about the benefit of a particular approach for textual description of differential privacy. For example, odds-based explanation methods from the study conducted by  Nanayakkara et al.~\cite{nanayakkara2023chances} were found to be effective in improving objective risk comprehension, subjective privacy understanding, and self-efficacy compared to examples-based methods. However, these approaches still leave room for improvement \cite{cummings2021need}.

Visual tools, while less commonly used, have shown considerable effectiveness in aiding understanding when used in conjunction with textual descriptions \cite{xiong2023exploring, nanayakkara2023chances}. For instance, Xiong et al.~\cite{xiong2023exploring} designed explanative illustrations that clarify how \(\varepsilon\) functions within DP models, leading to improved user comprehension. Moreover, Nanayakkara et al.~\cite{nanayakkara2023chances} explore effective methods of explaining epsilon \(\varepsilon\), particularly through visualization techniques that help users visualize the privacy-utility trade-offs. This body of work highlights the necessity for transparency in communicating epsilon \(\varepsilon\), suggesting that a well-informed public is more likely to make knowledgeable data-sharing decisions \cite{cummings2021need}.


\paragraph{Deployment and threat models.} Central Differential Privacy (CDP) and Local Differential Privacy (LDP) are two principal frameworks for implementing differential privacy. The two correspond to very different threat models from a security perspective, so effectively communicating these models is crucial.

Xiong et al. emphasize the importance of clear textual descriptions and visual aids to explain the role of the trusted curator and the benefits of minimal noise addition in CDP. Their research shows that visualizing the data flow from collection to anonymization and publication helps users understand how their data is protected at various stages \cite{xiong2022using}. Additionally, Xiong et al. found that descriptions focusing on the implications of DP rather than the technical definitions improved comprehension and willingness to share data \cite{xiong2020towards}.
%
Karegar et al. suggest using analogies and metaphors, such as adding noise to a signal before transmission, to help users grasp the concept of LDP. Interactive tools that allow users to experiment with data perturbation can demonstrate the effectiveness and implications of LDP, helping users visualize the privacy-utility trade-offs and understand the benefits of the model \cite{karegar2022exploring}.

Research by Xiong et al \cite{xiong2023exploring, xiong2022using} and Karegar et al \cite{karegar2022exploring} suggests that visual tools are particularly effective in explaining these models. Their work shows that when users are provided with clear illustrations that demonstrate the data flow and noise addition in CDP versus LDP, they better understand the implications of each model on their privacy and the utility of their data \cite{xiong2023exploring}. Additionally, Wen et al. used diagrams and pictures to illustrate the data perturbation process of Local DP to protect user privacy, further supporting the effectiveness of visual aids in communicating DP models \cite{wen2023influence}.


Existing studies do not go beyond LDP and CDP to more complicated threat models. Recent developments include the shuffle model~\cite{cheu2019distributed}, and the use of trusted hardware~\cite{maene2017hardware} or cryptography~\cite{pettai2015combining} to implement DP. The DP community has not settled on specific terminology for communicating these threat models, and comparing threat models remains challenging even for experts.

\section{Takeaways, Challenges, and Recommendations}
\label{sec:discussion}

This section summarizes the major takeaways of our review, with a particular focus on challenges and limitations identified by the studies we reviewed and our recommendations for addressing those challenges. Section~\ref{sec:chal_methodology} summarizes DP-specific takeaways associated with study methodology; Section~\ref{sec:chal_communication} summarizes takeaways from studies on communication about DP; and Section~\ref{sec:chal_tools} summarizes takeaways from studies on DP software tools.

\subsection{Methodology}
\label{sec:chal_methodology}
The DP usability studies in this SoK generally followed the best practices of human-computer interaction research methods. We believe the methodological challenges for future work are recruiting diverse DP users into research effectively, and increasing the validity and reliability in measuring DP usability. Below, we detailed these challenges our recommendations to overcome them.

\paragraph{Diversifying participant recruitment strategies.} For DP usability studies to achieve greater generalizability and to investigate deeper DP-related research questions, there is an increasing need to recruit both representative and specialized participants.
Assessing the effectiveness of DP communication often requires large samples from general populations, so many studies resorted to crowdwork platforms to quickly reach scalability and recruitment goals~\cite{bullek2017towards, cummings2021need, franzen2024communicating, franzen2022private, 10.1145/3576915.3623152, karegar2022exploring, kuhtreiber2022replication, nanayakkara2023chances, wen2023influence, xiong2020towards, xiong2022using, xiong2023exploring}. However, research has revealed systematic bias in crowdsourced data from these platforms~\cite{lee2018salience}. Even it is possible to control certain demographic variables to achieve a level of representativeness on platforms like Prolific, fully relying on crowdwork platforms for recruitment inherits the sample bias of the crowdworker population. Therefore, we recommend that future DP communication studies include more representative samples from the community in addition to crowdsourced samples to ensure the generalizability of research results.

On the other hand, in-depth DP usability studies, including those evaluating DP tools and DP communication among data practitioners, require participants to have certain technical and DP backgrounds. Existing studies had success recruiting from DP-specific listservs, job boards and professional networks ~\cite{gaboardi2018psi, marcusland2020, murtagh2018usable, nanayakkara2022visualizing, nanayakkara2024measure, ngong2024evaluating, sarathy2023don}, but samples tended to lack demographic or geographic diversity. 
Additionally, work-intensive studies that involve task completion exercises are resource-consuming, researchers had to proceed with smaller samples~\cite{garrido2022lessons, ngong2024evaluating}. To address the specific challenge to recruit specialized participants, we hope to see community efforts to establish a DP usability interest group consisting of representative groups of technical users with and without DP expertise who are interested in participating in future DP usability studies. The community can help in this process by supporting and publicizing resources for reaching DP stakeholders, such as the OpenDP mailing list and Slack channel.

Researchers in human-computer interaction acknowledged systematic bias of publishing user studies focusing on Western, Educated, Industrialized, Rich, and Democratic (WEIRD) participants~\cite{sturm2015weird}. The sample in current DP usability studies suffer from this bias. Growing the pool of potential study participants is the surest way to make experiments' subject more representative. DP research collaborations can extend their reach by broadening the geographic location, discipline and culture among project partners.

\paragraph{Accommodating different technical users.}
DP usability studies with specialized samples face the unique
challenges of assessing and managing participants’ prior technical
and DP expertise. The modern data professional has proficiency in either R or Python (specifically including the Pandas and Numpy packages), and those skills are prerequisite for working with DP tools. Pre-enrollment screening shows a baseline technical proficiency~\cite{alrige2013understanding,ngong2024evaluating}. Screening can also indicate a level for participants' DP knowledge~\cite{ngong2024evaluating}. can help assess participants' prior DP knowledge, which are necessary for more valid testing of DP tools. Convenience samples, without formal screening, are adequate for pilot testing but when they are applied to rigorous investigation of DP usability they are a missed opportunity to define the educational path toward DP literacy and competence. Our review shows that education materials (e.g., short paragraphs, handouts, tutorials, videos) used in studies~\cite{gaboardi2018psi, murtagh2018usable, ngong2024evaluating} help level the playing field of DP understanding among participants. We see an opportunity to establish standardized DP educational materials that articulate the path from technical proficiency for data analysis in R or Python to effective use of DP tools.

\paragraph{Improving validity through mixed-methods approach.}
Our SoK shows the advantages of the mixed-methods approach to examine the usability of DP. For example, Ngong et al.~\cite{ngong2024evaluating}'s usability study combined both quantitative usability metrics with qualitative data from surveys and interviews to generate deeper insights.
Similarly, studies incorporating both controlled experiments and post-study interviews articulated the difficulties users faced in interpreting DP parameters ~\cite{cummings2021need, nanayakkara2022visualizing, wen2023influence, xiong2020towards, xiong2022using, xiong2023exploring}. 
The mixed-methods approach allows for sophisticated study designs, but increase the difficulty to compare results from different studies.


\paragraph{Improving reliability by standardizing evaluation.}
Section~\ref{sec:method_evalaution} summarizes the varying constructs and metrics used in these studies to evaluate usability. Theses measures of users' understanding and satisfaction, of users' task performances and of users' self-described decision-making are tried and true human-computer interaction methodology. Even though the evaluation methods used in DP usability studies largely align with the human-computer interaction evaluation practices, they vary in terms of specific measurements and execution, making cross-study comparison difficult. We recommend that researchers in this area should develop standardized metrics and instruments that are specific to DP. For example, creating and validating 
a set of standardized multiple choice questions that cover a comprehensive range of DP knowledge (e.g. DP parameters, privacy budget, deployment model) will allow future studies to use a consistent metrics to measure participants' DP understanding.
This will increase measurements’ reliability and enable easier cross-study result comparison.

\subsection{Communication}
\label{sec:chal_communication}

A growing number of studies investigate communication strategies for DP, mostly with end users. We summarize the takeaways and remaining challenges in this area, and provide some recommendations for how to communicate DP guarantees to both technical users and end users.

\paragraph{Everyone struggles to understand DP parameters.}
Epsilon ($\varepsilon$) and other DP parameters define privacy guarantees, yet their meaning remains opaque and non-intuitive since they require users to reason about probabilistic guarantees. Unlike encryption, where higher bit lengths clearly indicate stronger security, DP parameters lack direct real-world analogies, making their implications difficult to grasp, \cite{cummings2021need, nanayakkara2023chances, franzen2022private}. Even participants with technical backgrounds struggle to interpret how changes in $\varepsilon$ impact privacy risk and data utility \cite{franzen2022private}.  Mathematical complexity further complicates communication. $\varepsilon$ and $\delta$ rely on probability theory and worst-case guarantees, concepts unfamiliar to most users \cite{xiong2023exploring}. Oversimplified explanations often distort or mislead privacy guarantees, while formal definitions remain inaccessible to end users~\cite{franzen2022private}. The privacy-utility trade-off is nonlinear and context-dependent, making it difficult to communicate effectively. Users often assume that lowering $\varepsilon$ always improves privacy, overlooking its dependence on dataset size, sensitivity, and query structure \cite{cummings2021need}.  

No single communication strategy works universally. Text-based explanations, visual aids, and risk-based descriptions each offer partial improvements, but usability of DP studies indicate that communicating DP parameters require tailored communication strategies or adapting explanations of DP parameters for diverse audiences \cite{xiong2023exploring}. For instance, technical users with DP expertise might prefer mathematical descriptions, but policymakers and data practitioners often require context-driven explanations. Furthermore, DP lacks standardized benchmarks for defining “safe” $\varepsilon$ or $\delta$ values, making it challenging for practitioners and policymakers to assess privacy guarantees \cite{franzen2022private}.   
Improving communication requires using accurate layman’s terms, contextual examples and relatable scenarios, and risk communication formats to explain DP parameters clearly \cite{nanayakkara2023chances}. Combining textual descriptions with visual aids enhances comprehension for non-technical users \cite{xiong2023exploring}. Standardized, intuitive frameworks for explaining DP parameters and trade-offs are necessary to support better decision-making across stakeholders.

\paragraph{End users struggle to understand deployment models.}
Communicating deployment models (e.g. central DP vs local DP) is challenging due to misinterpretation of trust assumptions. Users often assume that LDP always provides stronger privacy or misunderstand the role of a data curator in CDP \cite{nanayakkara2023chances}. The privacy-utility trade-off remains difficult to convey, as users struggle to grasp why LDP introduces more noise while CDP preserves higher data utility \cite{franzen2022private, xiong2023exploring}. Without clear benchmarks, these trade-offs remain unclear. Inconsistent messaging across tools and studies further complicates understanding, as many fail to explicitly clarify which model they use and its implications, leading to confusion among practitioners, policymakers, and end-users \cite{cummings2021need}. 

To improve communication, explanations of DP deployment models should be simplified with clear, accessible descriptions of the privacy-utility trade-offs and security implications of each model \cite{cummings2021need}. Visual aids, such as comparative illustrations and interactive tools, can enhance understanding of data flow and noise addition processes \cite{xiong2023exploring, karegar2022exploring, nanayakkara2022visualizing}. Contextual examples and real-world scenarios should be used to clarify operational nuances \cite{nanayakkara2023chances}. Metaphors and analogies can help make these abstract concepts and DP models more relatable, further aiding comprehension \cite{karegar2022exploring}. Standardized frameworks incorporating these approaches are necessary to support informed decision-making across diverse audiences.

\paragraph{End users struggle with text descriptions.}
To make communicating straightforward for the audience, text-based methods are popular as educational materials for differential privacy. However, simplifying complex DP concepts while maintaining accuracy remains difficult, as technical details often overwhelm users \cite{franzen2022private}. Balancing clarity and depth is essential, yet finding the right level of detail for diverse audiences is challenging \cite{xiong2022using}. Additionally, making textual explanations relatable and applicable across different user contexts requires tailored approaches, as users interpret privacy risks differently based on their background and expertise \cite{cummings2021need, nanayakkara2023chances, xiong2023exploring}. 
To improve comprehension, textual descriptions should use clear, layman-friendly language and integrate relatable narratives to illustrate key concepts \cite{franzen2022private}. Combining text with visual aids enhances understanding and reduces cognitive load \cite{xiong2022using}. Contextual examples that situate DP settings in familiar scenarios may help users grasp abstract concepts more intuitively \cite{cummings2021need, nanayakkara2023chances}.

\paragraph{Visual descriptions are more effective for end users.}
Non-textual methods, including visualizations, diagrams, and pictures, aim to simplify DP concepts, but several challenges persist. Ensuring accuracy without misinterpretation is difficult, as abstract DP principles do not always translate clearly into visual formats \cite{nanayakkara2023chances, xiong2022using}. Users often misinterpret visualizations without sufficient context, and engagement remains a challenge when designing intuitive and informative visuals \cite{karegar2022exploring}. Visual metaphors must balance simplicity and accuracy to avoid misleading interpretations of privacy guarantees \cite{xiong2023exploring}. 
Effective non-textual descriptions should integrate structured visual elements with clear textual explanations. Privacy-utility trade-off graphs, interactive models, and animations can help users explore different DP settings dynamically \cite{nanayakkara2023chances, nanayakkara2022visualizing}. Analogies and step-by-step diagrams can clarify DP mechanisms like data perturbation and aggregation \cite{xiong2023exploring, karegar2022exploring}. Iterative user testing should guide the refinement of these tools to ensure accessibility and accuracy across diverse audiences \cite{xiong2022using}.

\subsection{Software Tools}
\label{sec:chal_tools}
DP software tools are maturing quickly and have significantly improved in the past several years. Here, we summarize some of the main challenges still remaining for DP software tools, along with our recommendations.

\paragraph{Tools require DP expertise.}
Current tools---even those aimed at technical users without DP expertise---currently require significant DP expertise.
In particular, API tools require technical expertise for effective use~\cite{ngong2024evaluating}, and visual tools still demand an understanding of DP concepts~\cite{sarathy2023don}. We recommend that tool designers be conservative in their descriptions of expertise requirements, since deployments by technical users leveraging these tools may be incorrect. For tools that target technical users both with and without DP expertise, we recommend separate interfaces for different levels of DP expertise, employing progressive disclosure to reveal complexity only when necessary, and high-quality documentation with examples for all user levels.

\paragraph{Tools require manual parameter-setting.}
Current tools require manual setting of DP parameters, which means deep understanding of DP is required to use them~\cite{ngong2024evaluating, sarathy2023don, murtagh2018usable}. This can be incredibly challenging, even for DP experts: for example, even experts do not currently agree on how to set or interpret the privacy budget. Moreover, manual configuration of privacy parameters is complex and error-prone. We recommend adding automation for parameter-setting where possible, to reduce the burden on users. However, tool designers should be careful to avoid defaults that may violate DP, even if they improve automation.

\paragraph{Tools lack flexibility on deployment model and integration.}
Existing tools are designed for specific deployment models; they provide limited flexibility for altering the deployment model and are often unclear about the threat model they implement (for example, what level of trust in the data curator is required). We recommend that tool designers publish a clear description of each stakeholder's role and responsibilities, expected expertise, and expected level of trust. Moreover, current tools support only the central model of DP; we hope to see future tools support other deployment models.

Current tools provide limited integration with existing data processing pipelines and infrastructure. We recommend moving toward flexibility in integrating DP tools with non-DP data infrastructure, including modular design that allows experts to develop new extensions to integrate with existing data infrastructure.

\paragraph{Tools do not communicate utility.}
Current tools provide minimal support for understanding the utility of the DP results. Visual tools like DP Creator are beginning to investigate simple metrics for communicating expected utility to analysts, but API-based tools do not provide this support. We believe that tools should support negotiating the privacy-utility tradeoff among various stakeholders, by allowing the exploration of the impact on utility of various privacy parameter choices.

\paragraph{Some tools lack correctness checking.}
When using DP tools, technical users with and without DP expertise make mistakes that violate DP~\cite{ngong2024evaluating}. Automatic correctness checking can catch these mistakes, but reduces user satisfaction~\cite{ngong2024evaluating}. Many existing tools are research prototypes or aimed at users with DP expertise, and do not ensure correctness. However, since even DP experts routinely make mistakes when implementing DP deployments, we recommend that tools \emph{always} perform correctness checks, and raise errors when DP might be violated. Automated parameter-setting can help avoid DP mistakes, but automated approaches should cautiously avoid defaults that could silently violate DP. To improve user satisfaction, tools should follow existing APIs when possible, and perform correctness checking transparently.

\section{Open Questions} \label{sec:open_questions}

Our review identified many exciting avenues for future research that have not yet been addressed by prior work. In this section, we summarize open questions in all areas of DP usability.

\paragraph{Can end users understand DP?}
Understanding DP requires understanding complex mathematical concepts and threat models. Prior work has shown that communicating these aspects of a guarantee using textual descriptions is difficult, but that simplified explanations tend to be more effective than detailed ones. Additional research is needed to determine if simple textual descriptions can sufficiently communicate important features of a DP guarantee.

\paragraph{Can visualizations help users understand DP?}
Visual tools seem to help individuals understand DP guarantees, including both the mathematical guarantee itself~\cite{nanayakkara2023chances} and the surrounding deployment details~\cite{xiong2022using} (e.g. LDP vs CDP). Research is needed to determine whether visual tools are useful for technical users, and what level of objective understanding can these tools conveyed.

\paragraph{How should we communicate with other stakeholders?}
The vast majority of research on communicating about DP has focused on end-users. Additional research should focus on stakeholders other than end users---including developers (who need assistance setting parameters), downstream data users (who need to understand utility) and policymakers (who need to understand the strength of a guarantee). Recruiting from these groups is challenging, but this research will be an important step toward increased DP adoption and transparency in DP deployments.

\paragraph{How should we communicate parameters beyond $\epsilon$?}
The vast majority of studies with end-users have investigated methods for communicating about DP itself or the value of $\epsilon$, and have excluded other important parameters like the unit of privacy. These parameters can be just as important as $\epsilon$ in determining the strength of the privacy guarantee, but little research have tackled them.

\paragraph{How should we communicate threat models?}
Prior work demonstrates that threat models in DP deployments are difficult for end-users to understand~\cite{xiong2022using}, and many DP deployments do not fully document their threat models. We recommend that tools specify the threat model(s) they implement, and that DP deployments document the threat model in a standardized way---but how to develop and communicate the threat model of a DP deployment effectively to various stakeholders remains an open question.

\paragraph{How should we communicate about utility?}
We currently lack clear, standardized metrics to evaluate the privacy-utility trade-off, especially in domain-specific ways. We believe future studies should examine the effectiveness of utility and accuracy metrics for navigating the privacy-utility tradeoff, and also explore the use of visual tools for this purpose.

\paragraph{How should we measure understanding?}
Most of the studies in our review attempted to measure whether participants understood some concept (e.g. the definition of DP or the strength of the guarantee), but different studies used very different approaches to measuring understanding. Some evaluated understanding directly, by asking participants to reason about possible outcomes in order to demonstrate understanding; others evaluated understanding indirectly, by asking participants about their willingness to share data in a given situation. How best to measure participants' understanding of DP concepts, and whether standardized measures of understanding could help enable comparisons between studies, remains an important open question.

\paragraph{How do legal requirements influence DP deployments?}
Legal or regulatory requirements, such as Europe's General Data Protection Regulation (GDPR) and California's Consumer Privacy Act (CCPA), are important motivating factors for the deployment of DP in industry. As described in Section~\ref{sec:background}, a significant amount of work in the law community has sought to build connections between legal requirements and DP's formal guarantees. However, no studies have directly examined how legal requirements shape the use of DP and related techniques in practice. Garrido et al.~\cite{garrido2022lessons} found that ``\emph{[privacy practitioners] measure privacy based on the fulfillment of data protection regulation},'' suggesting that regulation is the \emph{primary} motivation for DP deployments in industry---but the study did not examine this question in detail, and we are not aware of any other studies that examine this issue. We recommend future research targeting practitioners involved in legal questions of privacy, to investigate the connection between the law and decisions to implement DP, and to provide guidance on how to consider legal requirements in the decision-making process for a DP deployment. 


\paragraph{Can users without DP expertise use DP tools effectively?}
Current research suggests that technical users without DP expertise struggle to use API-based tools~\cite{ngong2024evaluating}, but less research has investigated if visual tools can help technical users~\cite{sarathy2023don}. New wizard-based visual tools that combine guidance for parameter-setting with automation may enable technical users without DP expertise to correctly implement DP.

\paragraph{Can automation ease parameter-setting?}
Technical users with or without DP expertise struggle to set DP-specific parameters (e.g. clipping parameters) and choose the most appropriate mechanisms~\cite{ngong2024evaluating, sarathy2023don}. Some DP tools provide partial automation for setting some parameters, but no studies have investigated the impact of these features on usability.

\paragraph{Can verification tools be made accessible for all users?}
Software tools implement correctness checking in different ways, and existing work has shown that the specifics make a big difference in usability~\cite{ngong2024evaluating}. Future work should explore methods for DP verification tools that are usable by users regardless of their DP expertise.

\section{Conclusion}

In this paper, we systematically reviewed existing research on the usability of DP tools and DP communication. Our analysis highlights persistent difficulties in simplifying DP concepts, conveying the privacy-utility trade-off, inconsistent messaging and lack of standardized explanations and that current DP tools often require significant DP expertise making them less accessible to some technical users. To address these challenges, we recommend user-centered tool design, clearer communication strategies, and standardized messaging and educational resources  to improve accessibility and adoption. In addition, establishing a usability of DP interest group can also help in recruiting representative samples for usability of DP studies.

Future research should refine communication methods for DP parameters, investigate the effectiveness of visual aids for different user groups, and explore the intersection of DP practices with legal requirements.

\begin{acks}
This material is based upon work supported by the National Science Foundation under Grant No. 2238442 and 2336550, and by the Cold Regions Research and Engineering Laboratory (ERDC-CRREL) under Contract No. W913E521C0003. Any opinions, findings and conclusions or recommendations expressed in this material are those of the author(s) and do not necessarily reflect the views of the funding agencies.
\end{acks}

\bibliographystyle{plain}
\bibliography{refs}



\end{document}